\input{aipcheck}

\documentclass[
    ,final            
  ]
  {aipproc}

\layoutstyle{6x9}

\begin{document}

\title{FERO (Finding Extreme Relativistic Objects):  relativistic Iron K$\alpha$ lines in type 1 AGN.}


\author{A.L. Longinotti}{
  address={MIT Kavli Institute for Astrophysics and Space Research, Cambridge, Massachusetts (US)}
}

\author{I. de la Calle P\'erez}{
  address={European Space Astronomy Centre of ESA, Madrid (Spain)}
}


\begin{abstract}
The observational evidence for AGN relativistic Iron lines  is very much debated.           
To address this topic, the FERO project makes use of the largest sample of X-ray spectra of radio quiet Type 1 AGN 
available in the XMM-Newton archive. We perform a systematic fit of the individual sources using a full relativistic code.
Results on the first part of the project are presented here.

\end{abstract}

\maketitle


\section{The XMM-Newton FERO project}
The FERO project is based on one of the largest collection of Type-1 radio quiet AGN (149) ever assembled with targeted {\it XMM-Newton} observations. It aims at establishing on a statistical base how often the effects of X-ray illumination of relativistic accretion disks are present in AGN.
 In order to derive meaningful constraints on the properties of the (unknown) parent population of local radio-quiet AGN,  we applied the FERO selection criteria to the {\it RXTE} all-sky Slew Survey (\cite{rev}) and we selected the sources having a count rate
in the 3-8~keV energy band greater than 1 cts/sec.  This defines a flux-limited, almost complete (80\%) sample of 33 sources, 31 of which are included in the FERO analysis and provide an unbiased subsample of sources with good signal-to-noise ratio within the initial collection of 149 AGNs. 
The results of the spectral analysis (de la Calle et al. submitted) are presented here, whereas the study of the line profile in the stacked spectral residuals will be reported in Longinotti et al. (in prep). 

The spectra were fitted in the 2--10~keV band. The baseline model for the X-ray continuum comprises a power law, an intervening warm absorber and a Compton reflection component, plus 4 narrow Gaussian lines to model K$\alpha$ emission from Fe~I, Fe~XXV, Fe~XXVI and Fe~I K$\beta$, and a  Gaussian line at 6.3~keV with width set to 50~eV to model the Fe I Compton shoulder. The effects of general relativity are taken into account by analysing the data with the suite of routines  KY (\cite{dov}). Within this suite, {\it kyrline} has been specifically designed to constrain the parameters of the broad relativistic line. The spin of the black hole, the index of the emissivity profile of the accretion disk and the disk inclination angle to the observer are  free parameters. The intensity of the relativistic Fe line against the continuum (i.e. its Equivalent Width, EW) is used as a proxy for line detection.
The statistical study of the flux limited sample sources show that:

\begin{figure}[h!]
  \includegraphics[height=.44\textheight]{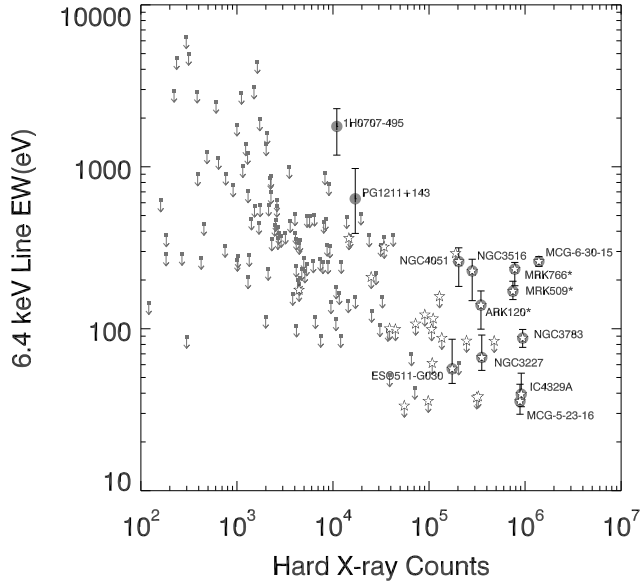}
\caption{The {\it XMM-Newton} FERO collection: broad Fe~K$\alpha$ line EW versus 2-10~keV counts. Filled circles mark 5$\sigma$ detections of the Fe line, stars mark the 31 sources in the flux limited sample.}
\end{figure}
\begin{itemize}

\item The fraction of sources in the FERO sample (Fig. 1)  that
present 5$\sigma$ detections for a relativistic Fe K$_\alpha$ line
is 9\% (13/149), but this fraction rises to 36$\%$ (11/31) for AGNs in the flux-limited sample.
Considering the upper limits to the line EW,  a broad line at the level of 40~eV can be rejected only in 4/31 objects. 
\item There is no significant difference between
Sy and QSO in terms of detection fraction. All detections have luminosities below $\sim$1 $\times$ 10$^{44}$ erg s$^{-1}$. 
\item The average relativistic Fe K$_\alpha$ line EW is of the order of 100~eV.
\item The average disk inclination angle is $<\theta>$=28$\pm$5$^\circ$, consistent with an intrinsic random distribution
  of inclination angles.
\item The index of the radial dependence of the disk emissivity ($\propto$ r$^{-\beta}$) is $<\beta>$=2.4$\pm$0.4, with a wide spread of values.
\item The spin value {\it a} is in general poorly constrained except for MCG-6-30-15 ({\it a}=0.86$^{+0.01}_{-0.02}$) and MRK509 ({\it a}=0.78$^{+0.03}_{-0.04}$)
\item The broad line EW is not  significantly dependent on the above line parameters.
\item No correlation was  found between the line EW or disk emissivity ($\beta$) and the source physical
  properties investigated, such as, black hole mass, accretion rate, 2--10~keV 
   luminosity and optical H$\beta$ FWHM.
\end{itemize}

\bibliographystyle{aipprocl}

\begin{thebibliography}{2}

\bibitem{rev}                          
Revnivtsev, M., Sazonov, S., Jahoda, K., \& Gilfanov, M. \emph{A\&A}, \textbf{418}, 927--936 (2004) 

\bibitem{dov}
Dov{\v c}iak, M., Karas, V., \& Yaqoob, T. \emph{ApJS}, \textbf{153}, 205--221 (2004)

\end{thebibliography}

\end{document}